\documentclass[pre,twocolumn,showpacs,floatfix,amsmath,amssymb]{revtex4}

\usepackage{graphicx}
\usepackage{dcolumn} % Align table columns on decimal point
\usepackage{color}
\usepackage{bm}      % Bold math

\begin{document}
\title{Intermittency in the isotropic component of helical and non-helical turbulent flows}

\author{L.N. Martin$^1$ and P.D. Mininni$^{1,2}$}
\affiliation{$^1$ Departamento de F\'\i sica, Facultad de Ciencias Exactas y
         Naturales, Universidad de Buenos Aires and CONICET, Ciudad 
         Universitaria, 1428 Buenos Aires, Argentina. \\
             $^2$ NCAR, P.O. Box 3000, Boulder, Colorado 80307-3000, U.S.A.}
\date{\today}

\begin{abstract}
We analyze the isotropic component of turbulent flows spanning a broad range of Reynolds numbers. The aim is to identify scaling laws and their Reynolds number dependence in flows under different mechanical forcings. To this end, we applied an SO(3) decomposition to data stemming from direct numerical simulations with spatial resolutions ranging from $64^3$ to $1024^3$ grid points, and studied the scaling of high order moments of the velocity field. The study was carried out for two different flows obtained forcing the system with a Taylor-Green vortex or the Arn'old-Beltrami-Childress flow. Our results indicate that helicity has no significant impact on the scaling exponents. Intermittency effects increase with the Reynolds number in the range of parameters studied, and are larger than what can be expected from several models in the literature. Finally, we confirm previous results showing the use of the SO(3) decomposition improves the convergence of scaling exponents.
\end{abstract}

\maketitle

\section{Introduction}
Turbulence is a recurring phenomenon in nature; we can find turbulent dynamics in atmospheric, geophysical and astrophysical flows. The dynamics of these flows is often described by identifying three characteristic ranges of scales \cite{frisch}: the injection range, whose properties depend on the forcing; the inertial range, which is assumed to have universal properties in the limit of infinite Reynolds number; and the viscous range, where dissipation takes place. In three dimensional isotropic and homogeneous turbulence, the scale separation between the forcing and the viscous range (i.e., the width of the inertial range) increases as a known power of the Reynolds number. Nowadays, computing power is scarcely sufficient to study in direct numerical simulations (DNS) flows with these three ranges well resolved. Even in the few simulations where an incipient scale separation is achieved \cite{goto,Kaneda,interacciones no locales}, an exploration of the parameter space to build confidence on such assumptions as universality of inertial range properties is currently out of reach.

In a recent study \cite{PRE1024} numerical simulations up to spatial resolutions of $1024^3$ grid points were performed using different forcing functions, including coherent and delta-correlated in time forcing, as well as using mechanisms that injected only energy or both energy and helicity into the flow. From numerical simulations \cite{borue,helicidad1,helicidad2,gomez} it is known that helical and non-helical isotropic and homogeneous turbulence follows a Kolmogorov scaling in the inertial range, albeit intermittency corrections. However, in \cite{PRE1024} a departure in the scaling exponents of higher order moments between helical and non-helical simulations was found. It was unclear whether these departures were associated to a dependence with the helicity content of the flow, or with anisotropies generated by the different forcing functions acting on the large scale. Other studies reported a dependence of the intermittency of the energy flux \cite{helicidad1} or of the recovery of isotropy on helicity-dependent statistical quantities \cite{helicidad3} on the amplitude of the helicity flux.

In this work, we use the SO(3) decomposition \cite{so1,SO(3)} (see also \cite{granrefer.SO(3)} for a review) to separate the isotropic and anisotropic components of a turbulent flow. To deal with the size of our datasets, the decomposition is implemented numerically as described in \cite{1}. Previous studies focused on the scaling of the anisotropic sectors (see e.g., \cite{so2,so4,so3}) from data stemming from numerical simulations or experiments, as it has been conjectured that both the isotropic and anisotropic sectors of turbulent flows may follow universal scaling laws in the inertial range (see e.g., \cite{so1,so5}), and as for the third order structure function it has been rigorously shown that the ``four-fifth law'' holds in the isotropic sector even in the presence of anisotropies \cite{SO(3),4/5aniso}. These studies also showed (albeit in simulations or experiments at moderate Reynolds numbers) that the use of the SO(3) decomposition improves the convergence of high order moments of the velocity field \cite{so1,so4}. For the determination of just the third order scaling exponent, studies of the isotropic component has been done up to spatial resolutions of $512^3$ grid points \cite{1}. However, less attention has been paid to comparisons of the scaling exponents in the isotropic sector for orders other than second or third, and to comparisons of these exponents for different forcing functions in simulations at large spatial resolution. Considering the differences reported in helical and non-helical turbulence, it seems reasonable to consider such flows for a comparison. Here, we use velocity fields stemming from six DNS with spatial resolutions ranging from $64^3$ to $1024^3$ grid points, using either helical or non-helical forcing mechanisms. Velocity increments and scaling exponents in the inertial range are computed for all runs, up to eight order for the simulations with the larger spatial resolution.

The analysis indicates that helicity has no measurable effect on the scaling exponents of the velocity field, with the differences observed in the exponents stemming from the different runs being within error bars, or ascribable to Reynolds number dependence. Moreover, we find that intermittency (as measured from the departure of the linear dependence of the scaling exponents with the order, associated to the development of strong events in the velocity field) increase with the Reynolds number in the range of parameters studied, indicating convergence of high order moments to their infinite Reynolds number limit is slow and may have not been achieved even in the largest numerical simulations up to date. Finally, we confirm previous results showing that the use of the SO(3) decomposition improves scaling laws and the determination of scaling exponents.

The structure of the paper is as follows. Section \ref{sec:simulations} briefly describes the numerical simulations. Section \ref{sec:SO3} introduces the SO(3) decomposition and describes the method we use to apply the decomposition to our data, based on the implementation proposed in \cite{1}. Section \ref{sec:scaling} discusses the $4/5$ law and the energy scaling in the isotropic sector, and Sect. \ref{sec:intermittency} presents the scaling laws obtained in the inertial range for the different moments of the velocity field. Also, a comparison between helical and no-helical flows is done, as well as a comparison with models of intermittency. Finally, Sect. \ref{sec:conclusions} presents the conclusions.

\section{Numerical simulations\label{sec:simulations}}

The datasets we use for the analysis stem from DNS that solve the momentum equation for an incompressible fluid with constant mass density. The Navier-Stokes equations under these conditions read
\begin{eqnarray} 
\frac{\partial \bf{v}}{\partial t} +\bf{v}\cdot\nabla \bf{v}&=&-\frac{\nabla p}{\rho}+\bf{f} + \nu \nabla^{2}\bf{v} \label{flujo}, \\
\nabla \cdot \bf{v}&=&0 \label{incompresible},
\end{eqnarray}
where $\bf{v}$ is the velocity field, $p$ is the pressure, $\bf{f}$ is an external force that drives the turbulence, $\nu$ is the kinematic viscosity, and $\rho$ is the mass density of the fluid (set to unity here in dimensionless units). The mode with the largest wave vector in the Fourier transform of  $\bf{f}$ is defined as the forcing wavenumber $k_{f}$, with the forcing scale given by $L_f=2\pi / k_{f}$.

Equations (\ref{flujo}) and (\ref{incompresible}) are solved using a parallel pseudo-spectral code in a three-dimensional box of size $2\pi$ with periodic boundary conditions \cite{Gomez05a,Gomez05b}. We use three different spatial resolutions: $64^3$, $256^3$ and $1024^3$ grid points. The equations are evolved in time using a second order Runge-Kutta method, and the code uses the $2/3$ rule for dealiasing. Reynolds numbers quoted are based on the integral scale and defined as $\textrm{Re}=UL/\nu$, where $U=\langle v^{2}\rangle^{1/2}$ is the r.m.s. velocity and the integral scale $L$ is defined as
\begin{equation}
L=2\pi \frac{\int E(k)k^{-1}dk}{\int E(k)dk},
\end{equation}
with $E(k)$ the energy spectrum such that the total energy is $E= \int E(k)dk$.

We examine two different flows, generated by different volume forces $\bf{f}$ that are either non-helical or fully helical (we consider a forcing function fully helical when $\left< \bf{f} \cdot \nabla \times \bf{f}\right>$ is maximal and non-helical when $\left< \bf{f} \cdot \nabla \times \bf{f}\right>$ is zero, with the brackets denoting spatial average): the Taylor-Green (TG) vortex \cite{tg}, and the Arn'old-Beltrami-Childress (ABC) flow \cite{abc}. The former is non-helical, and the resulting flow has no net helicity, although spatially localized regions with positive or negative helicity develop. The latter is fully helical, and the resulting flow therefore has helicity (where the flow helicity is defined as $H=\left< \bf{u} \cdot \nabla \times \bf{u}\right>$).

When using the TG vortex as a forcing function, we prescribe $\bf{f}$ as
\begin{eqnarray} \label{tg}
{\bf f}_{TG}=f_{0}\left[\sin(k_{f}x)\cos(k_{f}y)\cos(k_{f}z)\hat{x}\nonumber \right. \\ \left.-\cos(k_{f}x) \sin(k_{f}y)\cos(k_{f}z)\hat{y}\right],
\end{eqnarray}
while the ABC forcing is given by
\begin{eqnarray} \label{abc}
{\bf f}_{ABC}=f_{0}\left\{ \left[ B\cos(k_{f}y)+C\sin(k_{f}z) \right] \hat{x} + \right. \nonumber \\ \left[ A\sin(k_{f}x)+C\cos(k_{f}z) \right] \hat{y} + \nonumber \\ \left. \left[ A\cos(k_{f}x)+B\sin(k_{f}y) \right] \hat{z} \right\} .
\end{eqnarray}
Here $f_{0}$ is the forcing amplitude, which was set to have in the turbulent steady state all runs with r.m.s. velocities near unity. For ABC forcing, the constants were chosen to be $A=0.9$, $B=1$, and $C=1.1$. Table ~\ref{simulaciones} shows the parameters for all the runs. More details about the runs, and a detailed analysis of energy spectra, fluxes, and energy transfer, can be found in \cite{prl,PRE1024}.

\begin{table}
\caption{\label{simulaciones}Parameters used in the simulations. $N$ is the linear resolution, $\bf{f}$ is the forcing (either TG or ABC), $k_f$ is the forcing wave number, $\nu$ is the kinematic viscosity, and $\textrm{Re}$ is the Reynolds number.}
\begin{ruledtabular}
\begin{tabular}{cccccc}
Run & $N$ & ${\bf f}$ & $k_f$ & $\nu$ & Re\\ \hline
T1 & 64   &  TG  &2 &$5\times 10^{-2}$  &40  \\ 
T2 & 256  &  TG  & 2  &$2\times 10^{-3}$  & 675   \\ 
T3 & 1024 &  TG  & 2  &$3\times 10^{-4}$  & 3950   \\
A1 & 64   &  ABC & 3 &$4\times 10^{-2}$  &70  \\ 
A2 & 256  &  ABC & 3 &$2\times 10^{-3}$  & 820   \\  
A3 & 1024 &  ABC & 3  &$2.5\times 10^{-4}$& 6200  \\
\end{tabular}
\end{ruledtabular}
\end{table}

\section{The SO(3) decomposition\label{sec:SO3}}

The isotropic component of the longitudinal velocity structure functions for each flow is extracted using the SO(3) decomposition following the method described in \cite{1}. Some modifications were made considering the size of our simulations and to obtain a parallel version compatible with the way data is distributed among computing nodes. In this section we briefly introduce the method described in \cite{1} and give details of our implementation.

To do the SO(3) decomposition and recover the isotropic component, the longitudinal velocity structure function of order $p$ is decomposed in terms of the spherical harmonics $Y_{lm}$, obtaining
\begin{eqnarray}
S_{p}(\bf{l})&=&\left\langle \left\{ \left[ \bf{v}(\bf{r})-\bf{v}(\bf{r}+\bf{l})\right]\cdot \hat{l}\right\}^{p}\right\rangle \nonumber\\
%=\left\langle \sum_{jm}S_{P}^{jm}\left(\bf{r},l\right)Y_{jm}
%(\hat{l})\right\rangle=\nonumber\\=
%\sum_{jm}\left\langle S_{P}^{jm}(\bf{r},l)\right\rangle Y_{jm}(\hat{l})
&=&\sum_{jm}S_{p}^{jm}(l)Y_{jm}(\hat{l}) ,
\end{eqnarray} 
where the brackets denote spatial average over the variable $\bf{r}$, and homogeneity is assumed. The coefficients $S_{p}^{jm}$ are obtained by projecting $S_p({\bf l})$ into the spherical harmonics, or equivalently, by projecting the $p$-th power of the longitudinal increments
\begin{equation}
\delta v ({\bf r},{\bf l}) = \left[ {\bf v}({\bf r})-{\bf v}({\bf r}+{\bf l}) \right] \cdot \hat{l}
\end{equation}
into the spherical harmonics and averaging over $\bf{r}$. We are interested in the isotropic sector, given by the $S_p^{00}(l)$ functions,
\begin{eqnarray} \label{i1}
S_{p}^{00}(l)=\frac{1}{4\pi} \left< \int_{0}^{2\pi}\int_{0}^{\pi} \delta v^p ({\bf r},{\bf l}) l^{2} \sin(\varphi) d\varphi d\theta \right> ,
\end{eqnarray}
where the angles $\varphi$ and $\theta$ are associated to the vector ${\bf l}$. The discrete version of this equation, and therefore the expression used in the numerical code to carry out the decomposition, is
\begin{equation}
S_{p}^{00}(l) = \frac{1}{N_{d} N^{3}}\sum_{j=1}^{N_{d}} \sum_{i=1}^{N^{3}}\delta v^p({\bf r}_{i},{\bf l}_{j}),
\label{S00}
\end{equation}
where $N$ is the linear resolution and $N_{d}$ is the number of directions used to compute the average over the sphere. 

In Ref. \cite{1} it was shown that 146 different directions ${\bf l}_{j}$ covering in an approximately uniform way the sphere can be generated on a regular grid in such a way that all integer multiples of ${\bf l}_{j}$ lie on a grid point. This avoids the need to use three dimensional interpolations to compute the longitudinal increments $\delta v({\bf r}_{i},{\bf l}_{j})$ when ${\bf r}+{\bf l}$ does not lie on a grid point, significantly reducing the computational cost of the decomposition. The 146 directions are generated by the vectors $(1,0,0)$, $(1,1,0)$, $(1,1,1)$, $(2,1,0)$, $(2,1,1)$, $(2,2,1)$, $(3,1,0)$, $(3,1,1)$ and those that are obtained by permuting their components in every possible way (including multiplication of the vectors by $-1$). With periodic boundary conditions, negative multiples of the vectors are not needed, since a spatial average over the entire box with the increment ${\bf l}_j$ gives the same result as the average with $-{\bf l}_j$. This reduces $N_d$ in Eq. (\ref{S00}) to 73.

The code we use to solve the Navier-Stokes equations is parallelized using a two-dimensional domain decomposition \cite{Gomez05a,Gomez05b}. Each computing node stores a slice of the velocity field in real space of size $N\times N\times N_z$ ($N_z\le N$, with $N_z$ a function of the number of computing nodes). As a result, increments in Eq. (\ref{S00}) in the $x$ and $y$ directions can be computed locally in each node. However, increments in the $z$ direction require communication which is handled using the Message Passing Interface (MPI) library. The sum in Eq. (\ref{S00}) is then computed as follows: for each increment ${\bf l}_{j}$, displacements of the velocity field in the $x$ and $y$ direction are computed. Communication is then performed to displace the velocity field in the $z$ direction if needed, and $\delta v({\bf r}_{i},{\bf l}_{j})$ is computed for all values of ${\bf r}_{i}$. Finally, the sum over all ${\bf r}_{i}$ is done. The process is repeated for integer multiples of ${\bf l}_{j}$ by just displacing the already displaced velocity field on ${\bf l}_{j}$ over and over again. In this way, all communications are done between nearest neighbors avoiding all-to-all communications.

\section{The $4/5$ law and the energy spectrum\label{sec:scaling}}
\begin{figure}
\includegraphics[width=8cm]{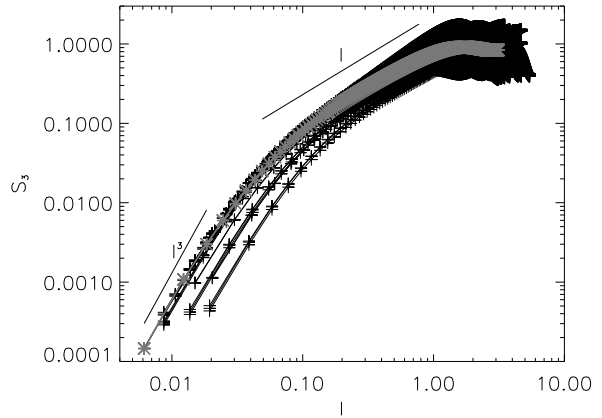}
\caption{\label{promedioTG1024}Third order structure functions $S_{3}({\bf l})$ and isotropic component $S_{3}^{00}(l)$ as function of $l$ for the T3 run. Black curves correspond to the 73 different directions while the gray curve is the $S_{3}^{00}(l)$ average. The straight lines indicate $\sim l^3$ and $\sim l$ scaling.}
\end{figure}

The result of computing the third order structure functions for the 73 directions using the T2 dataset is shown in Fig. \ref{promedioTG1024}. The average over all directions (computed using a procedure similar to the one described in \cite{1}) is also shown. From the $4/5$ law \cite{K1}, we expect the third order structure function to scale as the increment $l$ in the inertial range. This is indeed the case for the average (the isotropic component) in a wide range of scales, while the structure function in each direction may or may not follow this law. In the dissipative range, where the flow becomes regular, a scaling $\sim l^3$ is observed for all curves.

\begin{figure}
\includegraphics[width=8cm]{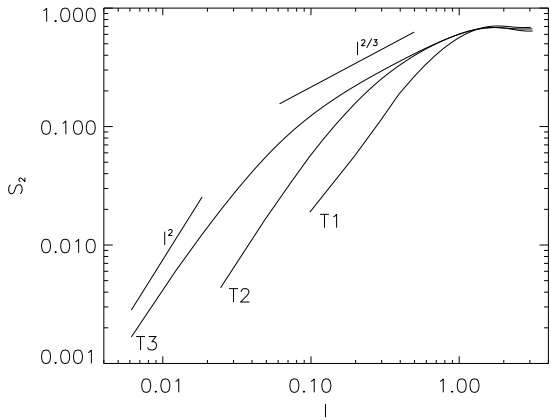}
\caption{\label{2}Isotropic component of the second order structure function for runs T1, T2, and T3 with increasing Reynolds number. The slopes of $2/3$ and $2$ are shown as references.}
\end{figure}

At lower Reynolds number (and spatial resolution) scaling in the inertial range can still be identified although the width of the scaling range decreases. This is illustrated in Fig. \ref{2} with the isotropic second order structure function $S_2^{00}(l)$ for runs T1, T2, and T3. The scaling in the inertial range of this function is associated with the scaling in the same range of scales of the isotropic energy spectrum. According to Kolmogorov theory (hereafter, K41) \cite{K1}, the second order structure function scales as $\sim l^{\zeta_2}$ with $\zeta_2 = 2/3$, which implies in turn an energy spectrum $E(k)\sim  k^{-(\zeta_2+1)} = k^{-5/3}$. The slope of these structure functions in the inertial range (the range where $S_3^{00} (l) \sim l$) is slightly larger than $2/3$, an effect associated to intermittency as discussed in more detail in the next section. As an example, a best fit to the power law in the second order structure function of run T3 gives $\zeta_{2}=0.702\pm 0.004$. Remarkably, at lower resolutions we also recover deviations from the K41 prediction for $\zeta_2$ and the energy spectrum, with run T2 giving $\zeta_{2}=0.69\pm 0.01$ and run T1 giving $\zeta_{2}=0.68\pm 0.04$. The second order isotropic structure functions for the ABC runs follow in the inertial range similar power laws, with slopes $0.66\pm 0.04$, $0.695\pm 0.006$ and $0.703\pm 0.003$ respectively for the resolutions of $64^3$, $256^3$ y $1024^3$ grid points. The six values are within error bars. At large Reynolds numbers, $S_2^{00}(l)$ is close to $\sim l^{-0.7}$ and the energy spectrum close to $E(k)\sim k^{-1.7}$ for both helical and non-helical cases, which is slightly steeper than $\sim k^{-5/3}$. This result is consistent with the result obtained from the largest simulation of isotropic and homogeneous turbulence done up to the moment using $4096^3$ grid points \cite{Kaneda,kaneda2}, and with other simulations at large spatial resolution and Reynolds number (see e.g., \cite{interacciones no locales}).

\section{Intermittency\label{sec:intermittency}}
\subsection{High order moments and anomalous scaling}

\begin{figure}
\includegraphics[width=8cm]{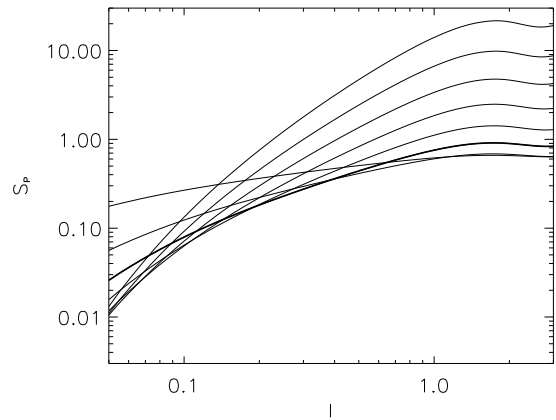}
\includegraphics[width=8cm]{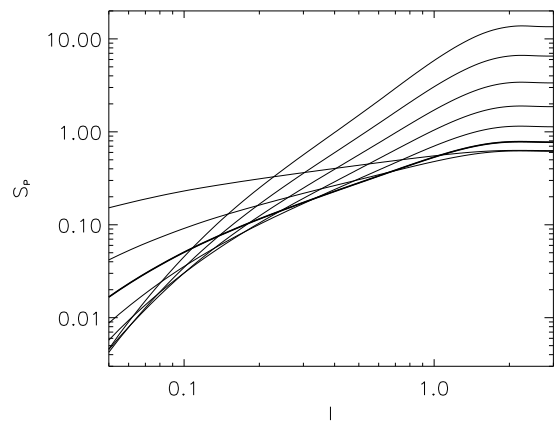}
\caption{\label{estructuralog-logTG1024} Above: $S_{p}^{00}(l)$ structure functions as function of the increment $l$ for $p$ from 1 to 8 for the T3 run. $S_{3}^{00}(l)$ is indicated by the thick curve. Only a range of scales near the inertial range is shown. Below: same for the isotropic structure functions for run A3.}
\end{figure}

Overall, the functions $S_{p}({\bf l})$ for all values of $p$ studied display, after averaging over all directions, a dissipative range that goes as $\sim l^p$, an inertial range following some power law, and a range at large scales that depends on the forcing. Each individual direction behaves as the average, although the structure functions for each direction show larger fluctuations and dispersion, specially at large scales. This can be understood in terms of anisotropies associated to the forcing which prevail at large scales. The differences between particular directions and the isotropic component decrease as the Reynolds number increases.

According to K41 theory, the longitudinal structure functions of order $p$ should scale in the inertial range as $\sim l^{\zeta_p}$ with $\zeta_p=p/3$, where $\zeta_p$ are the scaling exponents. This scaling corresponds to a scale invariant (non-intermittent) flow. However, turbulence comes in gusts and regions with strong gradients do not fill the entire space. Strong events are localized in space and time, and the probability of finding strong gradients at small scales is larger than what can be expected from a Gaussian distribution. Intermittency leads to anomalous scaling (deviations from the $\zeta_p = p/3$ relation), as the higher the order the more important are the contributions from these strong events. For the energy spectrum, the scarcity of the small scales is responsible for the steeper than $k^{-5/3}$ energy spectrum discussed in the previous section.

The dependence of the scaling exponent with the order is illustrated in Fig. \ref{estructuralog-logTG1024}. This figure shows the isotropic structure functions from $p=1$ to $8$ for runs T3 and A3. Only a range of scales near the inertial range is shown. In the inertial range, the longitudinal structure functions show scaling laws of the form $S_p^{00}(l) \sim l^{\zeta_p}$ as expected. As $p$ increases, the scaling exponent $\zeta_p$ increases monotonically. To check convergence of the structure functions and exponents, we computed the accumulated moments \cite{goto,interacciones no locales}, and verified that there is convergence up to the eight-order for these two runs (see e.g., \cite{interacciones no locales} for the accumulated moments in run T3 for the average in only two particular directions). For the runs with lower resolution, we computed structure functions up to the order $p$ according to the convergence of their accumulated moments.

\begin{figure}
\includegraphics[width=8cm]{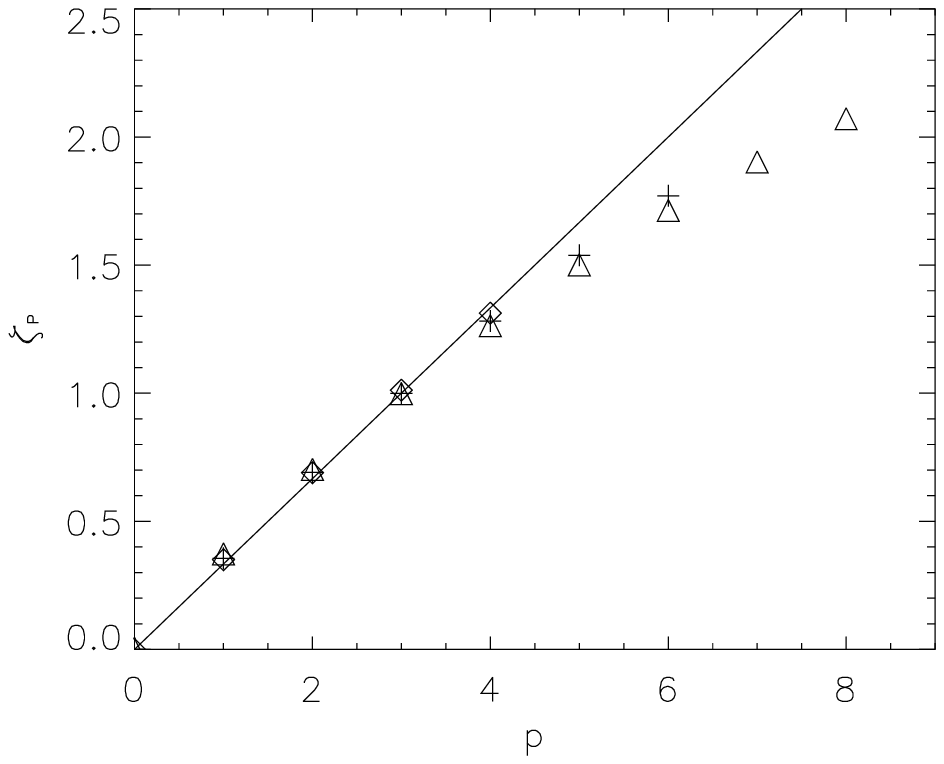}
\includegraphics[width=8cm]{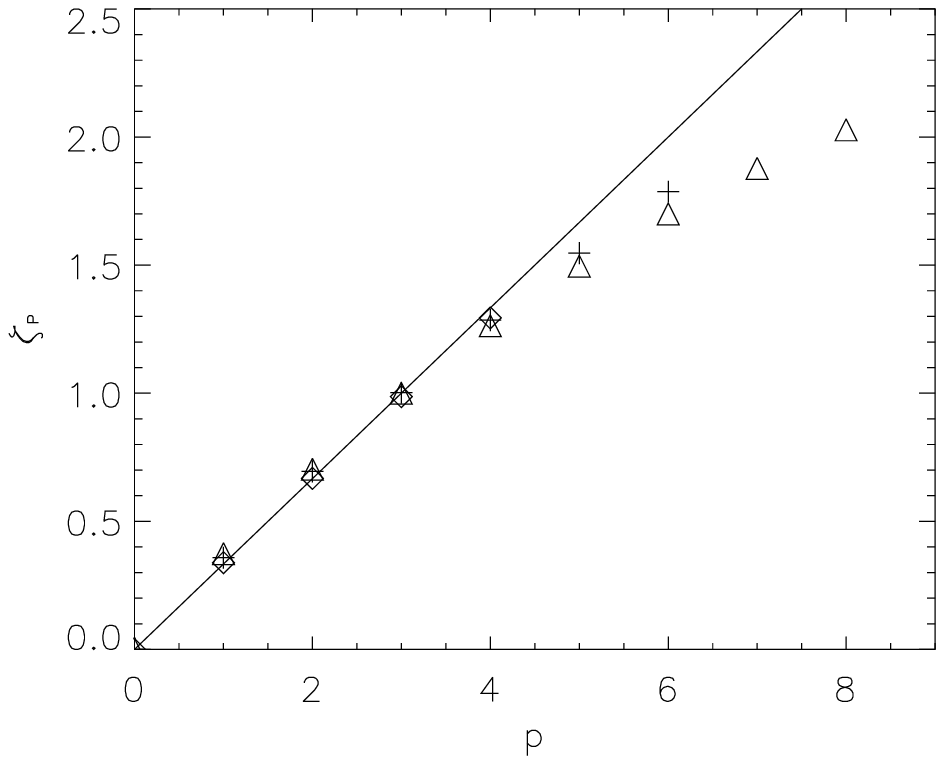}
\caption{\label{TG}Above: scaling exponents $\zeta_p$ in the T3 (triangles), T2 (crosses) and T1 runs (diamonds). The K41 prediction is given as reference. Below: same exponents for the A3 (triangles), A2 (crosses) and A1 runs (diamonds).}
\end{figure}

Let's now consider the scaling exponents $\zeta_p$ in the inertial range for all runs (where we define the inertial range as the range of scales where the 4/5 law holds). We start comparing the dependence of these exponents with the Reynolds number for the non-helical runs. Figure \ref{TG} shows $\zeta_p$ from the isotropic structure functions for the three runs with TG forcing. The $\zeta_p = p/3$ linear relation is indicated by the straight line. As previously mentioned, deviations of the exponents from the straight line are an indication of intermittency. Remarkably, the scaling exponents for the runs at lower Reynolds number are closer to the straight line than the exponents of the run at the largest Reynolds number, indicating intermittency still increases with the scale separation in the range of Reynolds numbers studied, and that more spatial resolution is required to reach convergence with Reynolds number for high order moments. Figure \ref{TG} also shows the same results for the ABC runs. For this flow, a similar dependence of the exponents with the Reynolds number can be identified. 

\begin{figure}
\includegraphics[width=8cm]{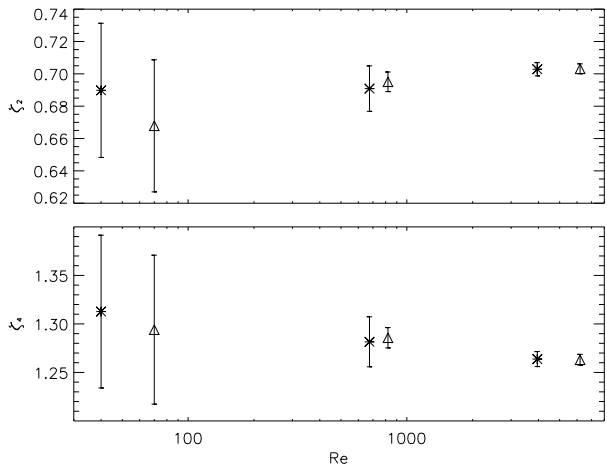}
\caption{\label{2y6}Second order (above) and fourth order (below) scaling exponents as a function of the Reynolds number for all runs (triangles are for ABC forcing and crosses for TG forcing).}
\end{figure}

\begin{figure}
\includegraphics[width=8cm]{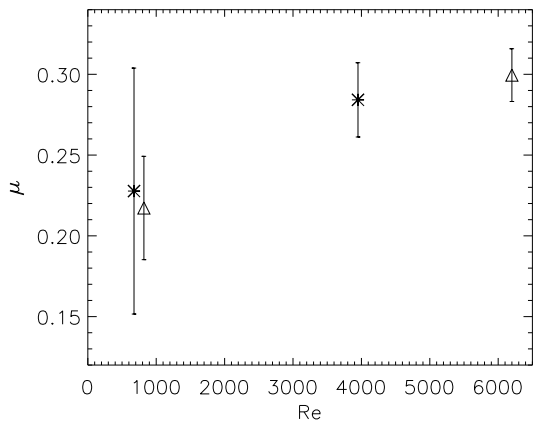}
\caption{\label{intermitencia}Intermittency exponent as a function of the Reynolds number for the four runs at larger Reynolds number (labels as in Fig. \ref{2y6}).}
\end{figure}

Figure \ref{2y6} shows the dependence of the second and fourth order scaling exponents for all runs as a function of the Reynolds number. Because of the different forcing functions used, even when comparing runs at the same resolutions the Reynolds numbers are slightly different. The second order exponent $\zeta_2$ slowly grows with Re, and seems to saturate for the largest Reynolds numbers reached. Note however that all values are within error bars. For $\zeta_4$ a decrease with the Reynolds number is observed, and in this case not all values are within error bars. The increase of the intermittency with Reynolds number is more pronounced when the intermittency exponent $\mu = 2\zeta_3-\zeta_6$ is studied as a function of Re (see Fig. \ref{intermitencia}; only the runs at larger Reynolds numbers are shown, as determination of $\mu$ requires the sixth order exponent). For the runs at the larger spatial resolution we obtain $\mu=0.28\pm 0.02$ and $\mu=0.30\pm 0.01$ respectively for T3 and A3, while for runs T2 and A2 we find $\mu= 0.23\pm 0.07$ and $\mu= 0.22\pm 0.03$ respectively. Our values for the highest Reynolds numbers agree with measurements made by Meneveau and Sreenivasan \cite{meneveu2} where they obtained $\mu= 0.26\pm 0.03$ (see also \cite{mu}). Based in the results in Fig. \ref{intermitencia}, $\mu$ may be even larger for larger values of the Reynolds number.

\subsection{Role of Helicity}
\begin{figure}
\includegraphics[width=8cm]{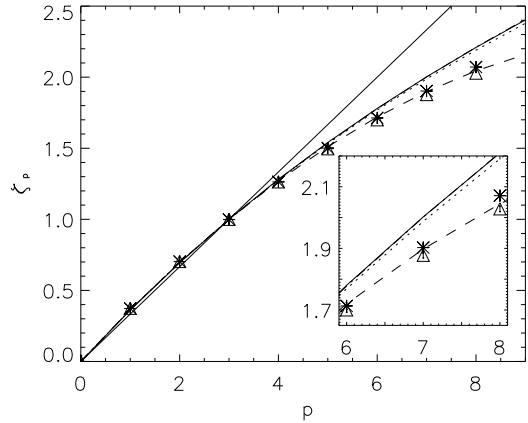}
\caption{\label{modelos}Comparison between the $\zeta_p$ exponents from the T3 (crosses) and A3 runs (triangles). Shown as a reference are the K41 (solid atraigth line), log-normal (dash), mean field (dotted), She-Leveque (dash-dotted), and Arimitsu and Arimitsu (solid) predictions, the mean field model and Arimitsu and Arimitsu model are indistinguishable. The inset shows a zoom for the highest order moments with error bars.}
\end{figure}

To study whether previously reported differences between helical and non-helical isotropic and homogeneous turbulence are associated to the presence of helicity or to anisotropies related to the forcing functions, we now focus on the datasets with the largest spatial resolution. Figure ~\ref{modelos} shows the scaling exponents as a function of $p$ up to the eight order for runs T3 and A3. The K41 prediction is shown as a reference, together with different models of intermittency: the log-normal model \cite{K62},
\begin{equation} \label{si}
\zeta_p=\frac{p}{3}+\frac{\mu}{18}(3p-p^2);
\end{equation}
where $\mu$ is the intermittency exponent previously defined, the She-Leveque model \cite{she-leveque},
\begin{equation}
\zeta_p=\frac{p}{9}+2\left[1-\left(\frac{2}{3}\right)^{p/3}\right];
\end{equation}
the mean-field approximation \cite{yakot}
\begin{equation}
\zeta_{p}=\frac{1.15}{3\left(1+0.05p\right)}p;
\end{equation}
and the model of Arimitsu and Arimitsu \cite{a-a,a-a2},
\begin{equation}
\zeta_{p}=\frac{\alpha_{0}p}{3}-\frac{2Xp^{2}}{9\left(1+C_{p/3}^{1/2} \right)} -\frac{1}{1-q}\left[1-\log_{2}\left(1+C_{p/3}^{1/2}\right)\right],
\end{equation}
where
\begin{equation}
C_{p/3}=1+2\left(\frac{p}{3}\right)^{2}\left(1-q\right)X\ln(2) ,
\end{equation}
and the quantities $\alpha_0$, $X$ and $q$ are determined from the intermittency exponent $\mu$ following the expressions in \cite{a-a,a-a2}.

The She-Leveque model and the mean field approximation have no free parameters, while the log-normal model and the model of Arimitsu and Arimitsu depend solely on the intermittency exponent. For the log-normal model, we use $\mu=0.28$ which is compatible with the value found in the A3 and T3 runs. Arimitsu and Arimitsu state that for infinite Reynolds number $\mu=0.22$ should be used in their model, and in \cite{a-a2} give explicit values for $\alpha_0$, $X$ and $q$ for this case. These are the values we use in Fig. \ref{modelos}, and when plotting the model with values corresponding to larger values of $\mu$ we observed an improvement in the model although measurable differences for the highest orders persist. The best fit to both helical and non-helical data is given by the log-normal model, although it is well known that for higher order moments the model will fail as its exponents do not increase monotonically with $p$. The data deviates from all the other models.

Table ~\ref{exp1024f} shows a comparison of the $\zeta_p$ exponents of the isotropic sector for the helical and non-helical runs at the largest spatial resolution. Except for the highest order computed, differences are within error bars, and we can thus conclude that previously reported differences measured without the SO(3) decomposition were associated to contributions from the anisotropic sector. The small discrepancy observed for $p=8$ may be related to Reynolds number dependence, as runs A3 has a slightly larger Reynolds number than run T3, and as it was noted before, the higher order exponents slowly decrease as the Reynolds number is increased.

\begin{table}
\caption{\label{exp1024f}Comparison of scaling exponents in helical and non-helical flows at the largest Reynolds number studied. $p$ is the order, $\zeta_{A3}$ are the isotropic scaling exponents of the helical run, $\zeta_{T3}$ are the exponents of the non-helical run, and $\zeta_{A3}-\zeta_{T3}$ is their difference.}
\begin{ruledtabular}
\begin{tabular}{cccc}
p & $\zeta_{T3}$ & $\zeta_{A3}$ & $\zeta_{A3}-\zeta_{T3}$ \\ \hline
1 & $0.372\pm 0.002$ & $0.373\pm 0.001$ & $0.001\pm 0.003$ \\
2 & $0.702\pm 0.004$ & $0.703\pm 0.003$ & $0.000\pm 0.007$ \\
3 & $0.998\pm 0.005$ & $0.999\pm 0.004$ & $0.00\pm 0.01$ \\
4 & $1.263\pm 0.007$ & $1.263\pm 0.005$ & $-0.00\pm 0.01$ \\
5 & $1.500\pm 0.009$ & $1.495\pm 0.006$ & $-0.00\pm 0.01$ \\
6 & $1.71\pm 0.01$ & $1.700\pm 0.007$ & $-0.01\pm 0.01$ \\
7 & $1.90\pm 0.01$ & $1.877\pm 0.008$ & $-0.02\pm 0.02$ \\
8 & $2.07\pm 0.01$ & $2.028\pm 0.009$ & $-0.04\pm 0.02$ \\
\end{tabular}
\end{ruledtabular}
\end{table}

\section{Conclusions\label{sec:conclusions}}

In this work we studied scaling exponents and intermittency in the isotropic sector of turbulent flows at different Reynolds numbers, generated by Taylor-Green forcing and by Arn'old-Beltrami-Childress forcing. The first is non-helical while the latter is maximally helical. The spatial resolutions were $64^3$, $256^3$ and $1024^3$ grid points, with the Reynolds number (based on the integral scale) ranging from $\approx 40$ to $\approx 6200$.

A dependence of the intermittency corrections with the Reynolds number was found, with a larger intermittency exponent as the Reynolds number was increased. When the SO(3) decomposition was used, no dependence of the level of intermittency with the helicity content in the flow was found, and small differences at the highest order studied may be due to Reynolds number dependence of the exponents. Finally, we confirm previous results indicating the SO(3) decomposition increases the range of scales where scaling is observed, and improves the determination of scaling laws.

When comparing with models of intermittency in the literature, it was found that the data is more intermittent than predictions from the She-Leveque model, the mean field approximation, and the model of Arimitsu and Arimitsu. In spite of the well known problems with the log-normal model, this model gives the best fit to our data. In this context, it is interesting to question the need to develop new models of intermittency without better data or a deeper understanding of the origin of intermittency in homogeneous and isotropic turbulent flows.

\begin{acknowledgments}
The authors acknowledge support from Grants No. UBACYT X468/08 and PICT-2007-02211. PDM acknowledges support from the Carrera del Investigador Cient\'{\i}fico of CONICET.
\end{acknowledgments}

\end{document}